\documentstyle[preprint,aps]{revtex}
\begin{document}
\draft
\title{Turbulence and finance?}

\maketitle

SIR: ~~Analogies between the price dynamics in the foreign exchange
market and 3-dimensional fully developed turbulence were recently
presented \cite{Ghashghaie96}.  Independently, we have carried out a
parallel study comparing the parallel of the dynamical properties of the
S\&P 500 index and of the time evolution of a 3-dimensional fully
turbulent fluid, but our study arrives at rather different
conclusions. Specifically, we find while intermittency -- i.e. abrupt
changes of activity in the time evolution of the variance of price
changes and of the mean energy dissipation -- and non-Gaussian behavior
(for short times) in the probability distribution of price and velocity
changes characterize both systems, the stochastic nature of the two
processes is {\it quantitatively\/} quite different. Among the
differences we find are:

\begin{itemize}
\item[(i)] {\it Price\/} changes of the S\&P 500 are substantially
uncorrelated (we detect only a slight enhanced diffusion), but {\it
velocity\/} changes in 3-dimensional turbulence are
anti-correlated. Specifically, we find that the measured autocorrelation
function of the price changes is a fast-decaying monotonic function with
a characteristic time of few minutes and the spectral density of the
index is a power-law $S(f) \propto f^{-2}$ for more than 4 orders of
magnitude, but in 3-dimensional turbulence an inertial range where $S(f)
\propto f^{-5/3}$ exists (Fig.1).

\item[(ii)] The maximum of the probability distribution $P(Z=0)$ of
price changes $Z_{\Delta t}(t)$ after a time $\Delta t$ shows clear
L\'evy (non-Gaussian) scaling (for $1\le \Delta t \le 1000$ minutes) as
a function of $\Delta t$ \cite{Mantegna95}, but the corresponding
turbulence quantity {\it does not\/} show scaling (Fig.~2).
\end{itemize}

One interesting result presented in \cite{Ghashghaie96} is that the
distribution of price changes in the foreign exchange market is changing
shape as a function of the time delay $\Delta t$.  The distribution
evolves from a leptokurtic (i.e. with tails fatter than in a Gaussian
distribution) to a Gaussian shape. Although a similar scenario is
observed in fully developed turbulence, the deep difference observed in
the correlation (of price and velocity changes) and scaling (of the
central part of the respective distributions) properties of the two
processes lead us to conclude that indeed, at the moment, the simplest
model describing the major features of the dynamics of prices in
speculative markets is the {\it truncated L\'evy flight} (TLF)
introduced in \cite{Mantegna95,Mantegna94,Mantegna-Nice}. In this model,
independent identically distributed price changes are characterized by a
L\'evy stable distribution in the central part of the distribution, but
a cut-off is present after which the distribution is not power-law. This
truncation implies that the process is characterized by a finite
variance. A TLF is only approximately self-similar and the process
eventually converges to a Gaussian distributed process due to the
finiteness of the variance, as predicted by the central limit
theorem. The TLF also describes quite well the dynamics of a price in
foreign exchange markets \cite{Arneodocm}.  Features observed in
economic data that are not explained in terms of the TLF model are (i)
the time dependence of the scale factor parameter $\gamma$ of the TLF
distribution, which shows a fluctuating behavior with bursts of activity
localized in limited intervals of time, and (ii) the long-range
correlations observed in the time evolution of $\gamma$ and in the
variance of price changes (two quantities related to the what is called
``volatility" in the economics literature) \cite{Mantegnaun}.

Can we reconcile the known intuitive parallels between finance and
turbulence with the fact that quantitatively the two phenomena are quite
different? This quantitative difference might disappear for a turbulent
system in an abstract space of non-integral dimensionality. Theoretical
studies \cite{Frisch76,Fournier78} (we thank A. Vulpiani for suggesting
these references) show that for $d\approx 2.05$ there exists (under the
Taylor hypothesis) an uncorrelated turbulent behavior characterized by a
spectral density $S(f)\propto f^{-2}$, just as for the S\&P 500. Further
study is required to test if a 2.05-dimensional turbulence could be
really consistent with the stochastic properties of market data and with
the main assumption of mathematical finance that is that no arbitrage is
possible in an efficient market \cite{Samuelson65}.

\begin{figure}
\caption{(a) Standard deviation $\sigma_Z(\Delta t)$ of the probability
distribution $P(Z)$ characterizing the price changes $Z_{\Delta t}(t)$
plotted double logarithmically as a function of $\Delta t$ for the S\&P
500 time series.  After a time interval of superdiffusive behavior
($0<\Delta t \le 15$ minutes), a diffusive behavior close to the one
expected for a random process with independent identically-distributed
increments is observed; the measured diffusion exponent 0.53 is very
close to the theoretical ({\it uncorrelated\/}) value 1/2.  (b) Standard
deviation $\sigma_U(\Delta t)$ of the probability distribution $P(U)$
characterizing the velocity changes $U_{\Delta t}(t)$ plotted double
logarithmically as a function of $\Delta t$ for the velocity difference
time series in turbulence (experimental data were kindly provided to us 
by Prof. K.R. Sreenivasan). Data recorded in the atmosphere at a Taylor
microscale Reynolds number $R_\lambda$ of the order of 1500.  After a
time interval of superdiffusive behavior ($0<\Delta t \le 10$), a
diffusive behavior close to the one expected for a fluid in the inertial
range is observed (the measured diffusion exponent 0.33 is close to the
theoretical ({\it anti-correlated\/}) value 1/3). (c) Spectral density of 
the S\&P 500 time series for the time period 1984--1987 representative 
of the 6-year time period 1984--1989 (an investigation performed for
the time period 1986--1989 gives a curve overlapping with the shown figure). 
The $1/f^2$
power-law behavior expected for a random process with independent
increments is observed over a frequency interval of more than 4 orders
of magnitude.  (d) Spectral density of the velocity time series of a
3-dimensional fully developed turbulent fluid. The $1/f^{5/3}$ inertial
range (low frequency) and the dissipative range (high frequency) are
clearly observed.}
\label{fig1}
\end{figure}
 
\begin{figure}
\caption{
(a) ``Probability of return to the origin'' $P(Z=0)$ for the S\&P 500
Index ($\circ$) and $P_g(Z=0)=1/ \sqrt{2\pi} \sigma (\Delta t)$ (filled
squares) as functions of the time sampling interval $\Delta
t$. $P_g(Z=0)$ is the probability of return to the origin expected for a
Gaussian stochastic process determined by measuring the standard
deviation $\sigma (\Delta t)$ of the experimental data. The two measured
quantities differ in the full interval implying that the profile of the
PDF must be {\it non-Gaussian}.  A power-law behavior is observed for the
entire time interval spanning three orders of magnitude. The slope of
the best linear fit is $-0.71\pm 0.025$. The difference between the two
quantities is decreasing when $\Delta t$ increases, implying a
convergence to a Gaussian process for high values of $\Delta t$. (b)
Probability of return to the origin $P(0)$ ($\circ$) and $P_g(0)$
(filled squares) (defined as in (a)) as functions of the time sampling
interval $\Delta t$ for the velocity of the fully turbulent fluid.
Again, the two measured quantities differ in the full interval, implying
that the profile of the PDF must be non-Gaussian.  However in this case,
a single scaling power-law behavior does not exist for the entire time
interval spanning three orders of magnitude.  The slope of the best
linear fit (which is of quite poor quality) is $-0.59 \pm 0.11$.}
\label{fig2}
\end{figure}

\author{Rosario N. Mantegna$^1$ and H. Eugene Stanley$^2$}

\address{$^1$ Istituto Nazionale di Fisica della Materia, Unit\`a di
Palermo and Dipartimento di Energetica ed
 Applicazioni di Fisica, Universit\`a di Palermo, Palermo, I-90128, ITALIA\\ 
$^2$Center for Polymer Studies and Dept. of Physics, Boston University,
Boston, MA 02215 USA}

\end{document}